\documentclass[review]{elsarticle}

\usepackage{hyperref}
\usepackage[T1]{fontenc}
\usepackage[hang]{subfigure}	
\usepackage{amsmath}
\journal{Journal of \LaTeX\ Templates}

\begin{document}

\begin{frontmatter}

\title{Commissioning and operation of the Cherenkov detector for proton Flux Measurement of the UA9 Experiment}

\author[mymainaddress]{F. M. Addesa\corref{mycorrespondingauthor}}
\cortext[mycorrespondingauthor]{corresponding author}
\ead{addesaf@roma1.infn.it}
\author[mythirdaddress]{L. Burmistrov}
\author[mythirdaddress]{D. Breton}
\author[mymainaddress]{G. Cavoto}
\author[mythirdaddress]{V. Chaumat}
\author[mythirdaddress]{S. Dubos}
\author[mysecondaryaddress]{L. Esposito}
\author[mysecondaryaddress]{F. Galluccio}
\author[mysecondaryaddress]{M. Garattini}
\author[mymainaddress]{F. Iacoangeli}
\author[mythirdaddress]{J. Maalmi}
\author[mysecondaryaddress]{D. Mirarchi}
\author[mythirdaddress]{A. Natochii}
\author[mythirdaddress]{V. Puill}
\author[mysecondaryaddress]{R. Rossi}
\author[mysecondaryaddress]{W. Scandale}
\author[mysecondaryaddress]{S. Montesano}
\author[mythirdaddress]{A. Stocchi}

\address[mymainaddress]{INFN - Istituto Nazionale di Fisica Nucleare - sezione di Roma, Piazzale A. Moro 2, Roma, Italy}
\address[mysecondaryaddress]{CERN, European Organization for Nuclear Research, Geneva 23, CH-1211 Switzerland }
\address[mythirdaddress]{LAL - Laboratoire de l'Accélérateur Linéaire -  Université Paris-Sud 11, Centre Scientifique d?Orsay,
B.P. 34, Orsay Cedex, F-91898 France }

\begin{abstract}
The UA9 Experiment at CERN-SPS investigates channeling processes in bent silicon crystals with the aim to manipulate hadron beams. 
Monitoring and characterization of channeled beams in the high energy accelerators environment ideally requires in-vacuum and radiation hard detectors. For this purpose the Cherenkov detector for proton Flux Measurement (CpFM) was designed and developed. It is based on thin fused silica bars in the beam pipe vacuum which intercept charged particles and generate Cherenkov light. The first version of the CpFM is installed since 2015 in the crystal-assisted collimation setup of the UA9 experiment.\\
In this paper the procedures to make the detector operational and fully integrated in the UA9 setup are described. The most important standard operations of the detector are presented. They have been used to commission and characterize the detector, providing moreover the measurement of the integrated channeled beam profile and several functionality tests as the determination of the crystal bending angle.
 The calibration has been performed with Lead (Pb) and Xenon (Xe) beams and the results are applied to the flux measurement discussed here in detail. 
\end{abstract}


\end{frontmatter}


\section{\textbf{Introduction}}

The primary goal of the UA9 experiment \cite{UA9} is to demonstrate the feasibility of a crystal-based halo collimation as a promising and better alternative to the standard multi-stage collimation system for high energy hadron machines. 
 \begin{figure}[htbp!!]
\begin{center}
\includegraphics*[width=0.9\textwidth]{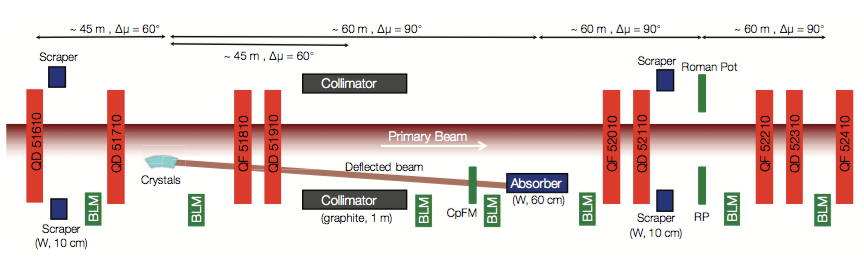}
\caption{UA9 experimental installation in LSS5 at CERN-SPS. The CpFM detector is located between the bent crystal and the absorber (about 60\,m from the crystal and $\simeq$4.5\,m from the absorber). Typical bending angles of SPS-UA9 crystals are $\simeq170\mu$\,rad. \cite{Montesano:IBIC2016-TUPG40}}
\label{ua9setup}
\end{center}
\end{figure}
The main installation of the experiment is located in the Long Straight Section 5 (LSS5) of the CERN Super Proton Synchrotron (SPS) and consists of a crystal-assisted collimation prototype. It is made by preexisting optical elements of the SPS and new installations including three goniometers to operate five different crystals used as primary collimators, one dedicated movable absorber, several scrapers, detectors and beam loss monitors (BLMs) to probe the interaction of the crystal with the beam halo \cite{MontesanoIPAC}. A schematic of the layout of the experiment is shown in Fig.\,\ref{ua9setup}.
The main process investigated is the so-called \textit{planar channeling}: particles impinging on a crystal having a small angle (less than $\theta_{c}$, called the critical angle for channeling) with respect to the lattice planes are forced to move between the crystal planes by the atomic potential. If the crystal is bent, the trapped particles follow the bending and are deflected correspondingly. When an optimized crystal intercepts the beam halo to act as collimator, about 80\% of the particles are channeled, coherently deflected and then dumped on the absorber (see Fig. 1), effectively reducing the beam losses in the sensitive areas of the accelerator \cite{1SCANDA,2SCANDA,3SCANDA,4SCANDA}.

In order to fully characterize this collimation system, it is essential to steadily monitor the flux of the halo particles deflected by the crystal towards the absorber. Typical crystal-extracted fluxes range from $10^5$ up to $10^7$ protons/s (i.e. from 1 up to 200 protons per SPS revolution) and about $10^5$ ions/s (1-3 ions per SPS revolution). Such a low flux does not allow to use the standard SPS instrumentation, for example BCTs (Beam Current Transformer \cite{Jones:1982418}) which are optimized for higher fluxes ($>10^9$\,protons/s). 
For this reason, the Cherenkov detector for proton Flux Measurement (CpFM) was designed and developed.\\
\section{\textbf{The Cherenkov detector for proton Flux measurement}}
The CpFM detector has been devised as an ultra-fast proton flux monitor. It has to provide measurements of the extracted beam directly inside the beam pipe vacuum, discriminating the signals coming from different proton bunches in case of multi-bunch beams, with a 25\,ns bunch spacing. It is also able to stand and to detect very low ion fluxes (1-3 ions per turn).
The sensitive part of the detector is located in the beam pipe vacuum in order to avoid the interaction of the protons with the vacuum-air interface, hence preserving the resolution on the flux measurement. All the design choices are explained in detail in \cite{1748-0221-12-04-P04029}.\\
  \begin{figure}[htbp]
\begin{center}
\includegraphics[scale=.3]{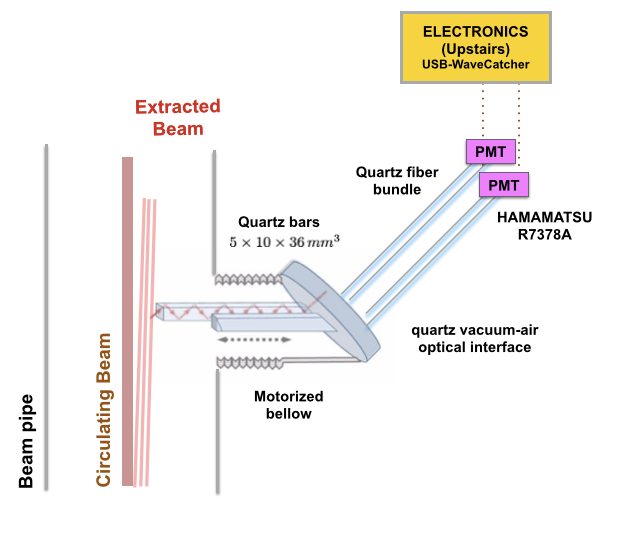}
\caption{Conceptual sketch of the CpFM in its first layout.}
\label{CpFM_final_layout}
\end{center}
\end{figure}
 A conceptual sketch of the first version of the CpFM is shown in Fig.\,\ref{CpFM_final_layout}. It consists of two identical Fused Silica bars ($5\times10\times360\,mm^3$, 5\,mm along the beam direction) acting as Cherenkov light radiators and light guides at the same time. When a relativistic charge particle crosses a bar, it produces Cherenkov light that is transported by internal reflection to the other tip of the bar. One bar is 5\,mm closer to the center of the circulating beam than the other one and is devoted to the flux measurement. The second bar, retracted from the beam, provides background measurements. The vacuum-air interface is realized by a standard quartz viewport. The light signal from each bar is guided onto a PMT (HAMAMATSU R7378, anode pulse rise time $\simeq 1.5\,ns$) by a 4 m long fused silica fibers bundle. The bars can gradually approach the extracted beam through a movable bellow on which the viewport is mounted.
  The PMTs are read-out by an ultra-fast analog memory, the 8-channels USB-WaveCatcher \cite{Breton}.
  The first prototype of the detector was calibrated at the Beam Test Facility (BTF) of Laboratori Nazionali di Frascati with 450\,MeV/c electrons and at the H8 external line of the North experimental area of CERN, with a 400 GeV proton beam \cite{1748-0221-12-04-P04029}. The relative resolution on the flux measurement of the CpFM for 100 incoming electrons was assessed to be 15\%, corresponding to a 0.63 photoelectron (ph.e.) yield per single particle \cite{1748-0221-12-04-P04029}. The CpFM is installed in the SPS tunnel since 2015. 
  \subsection{Electronic readout and DAQ system}
 The CpFM electronic readout is realized by the 8-channels USB-WaveCatcher board \cite{Breton}. This is a 12-bit 3.2\,GS/s digitizer; 
 6 other frequencies down to 0.4\,GS/s are also selectable via software.
Each input channel is equipped with a hit rate monitor based on its own discriminator and on two counters giving the number of times the programmed discriminator threshold is crossed (also during the dead time period corresponding to the analog to digital conversion process) and the time elapsed with a 1 MHz clock. This allows to measure the hit rate.
Each input channel is also equipped with a digital measurement block located in the front-end FPGA which permits extracting all the main features of the largest amplitude signal occurring in the digitizer window in real time (baseline, amplitude, charge, time of the edges 
with respect to the starting time of the acquisition).\\
 The WaveCatcher is triggered by the UA9 trigger (common to all the other UA9 instrumentation). This trigger signal is the SPS revolution signal (43 kHz) down-scaled by a factor of 1000 and synchronized with the passage of a filled bucket in LSS5. The acquisition rate corresponds to the trigger frequency (43 Hz). Three signals are acquired: two CpFM channels and the UA9 trigger itself.
 The board is equipped with a USB 2.0 interface for the data transfer. \\
 The off-line analysis used to characterized the CpFM signal and to perform the event identification \cite{Addesa:2661725} is mainly based on the output of the measurements block.

\section{Procedures preliminary to data taking: readout settings optimization}
Prior to every UA9 data taking a standard procedure is followed to prepare the detector for operation. It consists of checking the PMT gain stability and optimizing the gain of the PMTs and the settings of the WaveCatcher with respect to the characteristics of the beam to be measured. 
\subsection{PMT gain stability check}
\begin{figure}[htbp!!!]
\begin{center}
\includegraphics[scale=.25]{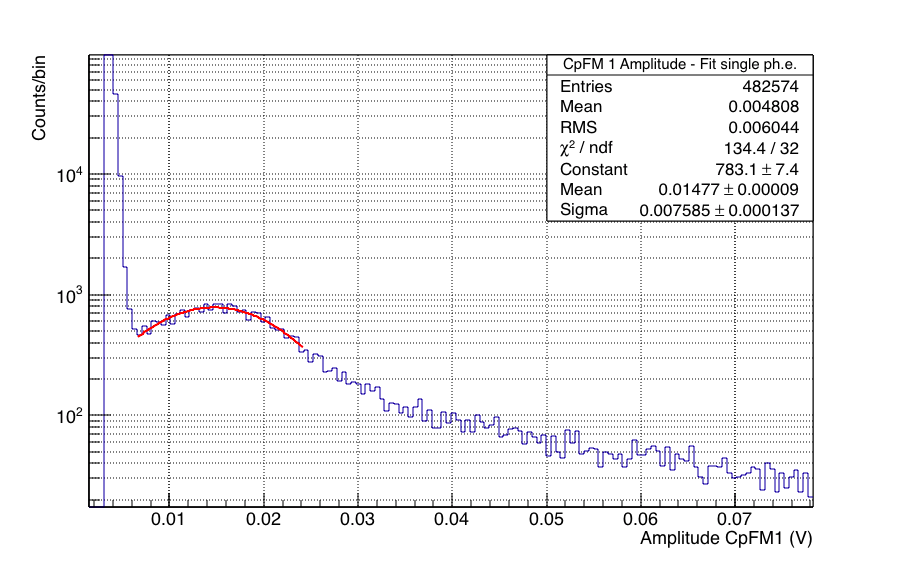}
\caption{CpFM channel 1 amplitude distributions when the detector is completely retracted from the beam pipe. This distribution is used to set the threshold for the rate measurement removing the electronic noise pedestal and also to check the gain stability of the PMT.}
\label{single_photoelectron}
\end{center}
\end{figure}

 The reliability of flux measurements depends on the stability of the calibration factor for which in turn the stability of the PMT performance is fundamental. For this reason, before every UA9 data taking, the stability of the PMT gain is checked through a simple procedure. It consists in a high-statistic ($10^5$) data acquisition of the CpFM signals when the detector is located at the parking position (10\,cm from the beam pipe center) and the beam in the SPS is already in coasting mode\footnote{In storage mode and with the RF cavity switched on in such a way the beam is kept bunched}. In this way, the amplitude distribution of the detector signals corresponds to the amplitude distribution of the background (Fig.\,\ref{single_photoelectron}), the latter being mostly composed by single photoelectron ($S_{ph.e}$) events plus a long tail due to particles showering by interacting with the aperture restrictions of the machine. If the PMTs are not affected by any gain variation, for example by radiation damage, the $S_{ph.e}$ position in the amplitude distribution is unchanged. 
\subsection{PMT gain optimization}
While choosing the PMT gain for both proton and ion runs, the maximum expected flux has to be considered together with the photoelectron yield per charge and the WaveCatcher dynamic range. To determine the optimal gain is noticed that the saturation of the electronics occurs at 2.5\,V.

The typical proton beam setup during UA9 experiments is a single 2\,ns long bunch of $1.15\times10^{11}$ protons stored in the machine at the energy of 270\,GeV \cite{JACoW-HB2012-TUO3A02}. 
 For this beam intensity, the beam flux deflected by the crystal toward the CpFM ranges from 1 up to $\simeq200$ protons per turn (every $\sim$23\,$\mu$s), depending on the aperture of the crystal with respect to the beam center. In this case the optimal PMT gain is $5\times 10^6$ corresponding to bias the PMT at 1050\,V. When the PMT is operated at such a gain one single ph.e. corresponds to $\sim$15\,mV (Fig.\,\ref{single_photoelectron}); considering the calibration factor (0.63 photoelectron yield per charge, measured at BTF and H8 line) the average amplitude of the signal produced by 200 protons is much lower than the dynamic range of the digitizer, allowing furthermore a safety margin of additionally 80 protons per pulse.\\

The typical ion beam setup during UA9 data taking consists in few bunches of $1.1\times10^{8}$ fully stripped Lead (Pb) or Xenon (Xe) ions \cite{JACoW-HB2012-TUO3A02}. As in the case of protons, the ion beam is in coasting mode at the energy of 270 GeV per charge. With such a beam intensity the ions to be measured by the CpFM per bunch and per turn is very low, from 0 to 3 ions. Nevertheless, since the Cherenkov light produced by a charged particle is proportional to the square of the charge of the particle
, few ions can be enough to saturate the dynamic range of the electronics. Therefore the PMT gain has to be 1 to 2 orders of magnitude smaller than the gain used for protons. \\The procedure followed for ions foresees to start with a bias voltage of 500\,V, corresponding to a gain of $\sim 2.5\times 10^4$. Then, depending on the expected flux, it can be increased in steps of 100\,V up to a maximum value of 700\,V. This bias corresponds to a gain of $\sim  2.5\times 10^5$, at which a flux of more than 1 ion per turn results, according to the calibration factor, in the saturation of the electronics. 
\subsection{WaveCatcher settings optimization}
In the following the optimal readout electronic settings are discussed with respect to the characteristics of the signal to be sampled.

\begin{figure}[htbp!!]
\centering
{\includegraphics[scale=.30]{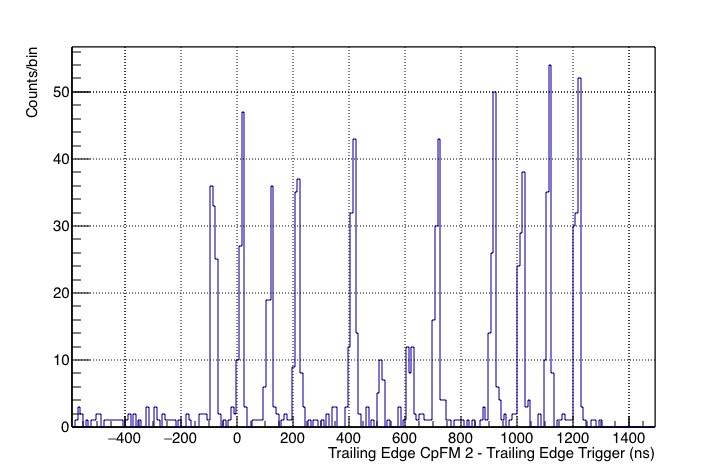}}
\caption{ Ion beam, November 23th, 2016. Bunch structure displayed by the arrival time distribution of the CpFM 2 signal with respect to the UA9 trigger signal. It consists in 3 trains of 4 bunches each. The bunch spacing is 100\,ns while the train spacing is 200\,ns. To display all the bunches the 2.4\,$\mu s$ digitizer window has been used ( sample frequency = 400\,MS/s.}
\label{bunch_structure}
\end{figure}

\textit{Sampling frequency and digitizer window lenght}. Since the PMT reading out the CpFM signal is very fast (rise time $\simeq$1.5\,ns), the highest sampling frequency, 3.2 GS/s, represents the best choice because it allows for a better reconstruction of the signal shape.\\ To use the 3.2\,GS/s sampling frequency a fine synchronization of the CpFM signals and the UA9 trigger is needed, the digitizer being started by the latter. 
The choice of the sampling frequency and therefore of the window length, defined as 1024 sample points divided by the sample frequency, is also influenced by the setup of the beam. 
For example, with an ion beam in multi-bunch mode it could be useful to first study all the bunches and then choosing to sample and to reconstruct more precisely only one of them (see Fig.\,\ref{bunch_structure}). In this case, first the 400\,MS/s sampling mode has to be selected in order to have an overview of all the bunches. Using then the 3.2\,GS/s mode and playing with the on-board trigger delay parameter, it is possible to center the window of the digitizer around the selected bunch. Moreover, having just one bunch in the digitizer window is essential to directly use 
the measurement block of the WaveCatcher.
 If more peaks are present in the same digitizer window, the measurement of the average parameters of the signal shape would be biased. 

\textit{Hit rate monitor threshold}. The hit rate monitor cannot be used to count the channeled particles because, if the beam is well bunched, they are deflected at the same time (or more precisely within the 2\,ns of the bunch), producing a single signal shape proportional to their number. Nevertheless, the hit rate monitor can be effectively used to quickly find the channeling orientation of the crystal or to align the CpFM with respect to the beam. In this case the CpFM has to detect only changes in the counts rate. The absolute value of the rate is not important and thus the threshold of the hit rate monitor can be kept just over the electronic noise, corresponding to the pedestal of the amplitude distribution of the background shown in Fig.\,\ref{single_photoelectron}.
\section{\textbf{An in-situ calibration strategy with ion beams}}
\label{calibration}
The SPS ion runs at the end of each year offer a possibility to calibrate in situ the detector. In fact in this case, the ph.e. yield per ion allows an excellent discrimination of the signal coming from 1, 2 or more ions.\\ 
 The Cherenkov light produced by a single ion of charge $Z$ is $Z^2$ times the light produced by a single proton. For example, as the charge of a completely stripped Lead (Pb) ion is 82, the light produced by a single ion is equal to 6724 times the light produced by a proton. During SPS ion runs for the UA9 experiment, each Pb ion charge is accelerated to 270 GeV, exactly as in UA9 proton beam runs.
 Identifying the amplitude signal corresponding to a single ion ($A_{Pb}$), the photoelectron yield per proton ($y$) can be obtained by: 
  \begin{equation}
 \label{recal}
 A_{Pb}=Z^2_{Pb}\times y\times S_{ph.e}\,(mV)
\end{equation}
where the $S_{ph.e}$\,(mV) depends on the PMT and it can be obtained fitting the amplitude distributions in Fig.\,\ref{single_photoelectron} and rescaling it to the PMT gain used for ions (700 V in this case).
The left side of the equation is provided by a data acquisition with both the bars intercepting the channeled beam. In this way the amplitude distribution of the channeled ions is easily obtained. This strategy has been applied for the first time during the Pb ion run in November 2016, providing reliable calibration factors for the flux measurement for those runs.
 \begin{figure}[htbp!!!]
\begin{center}
\subfigure[\label{ion_distribution}]{\includegraphics[width=0.7\textwidth]{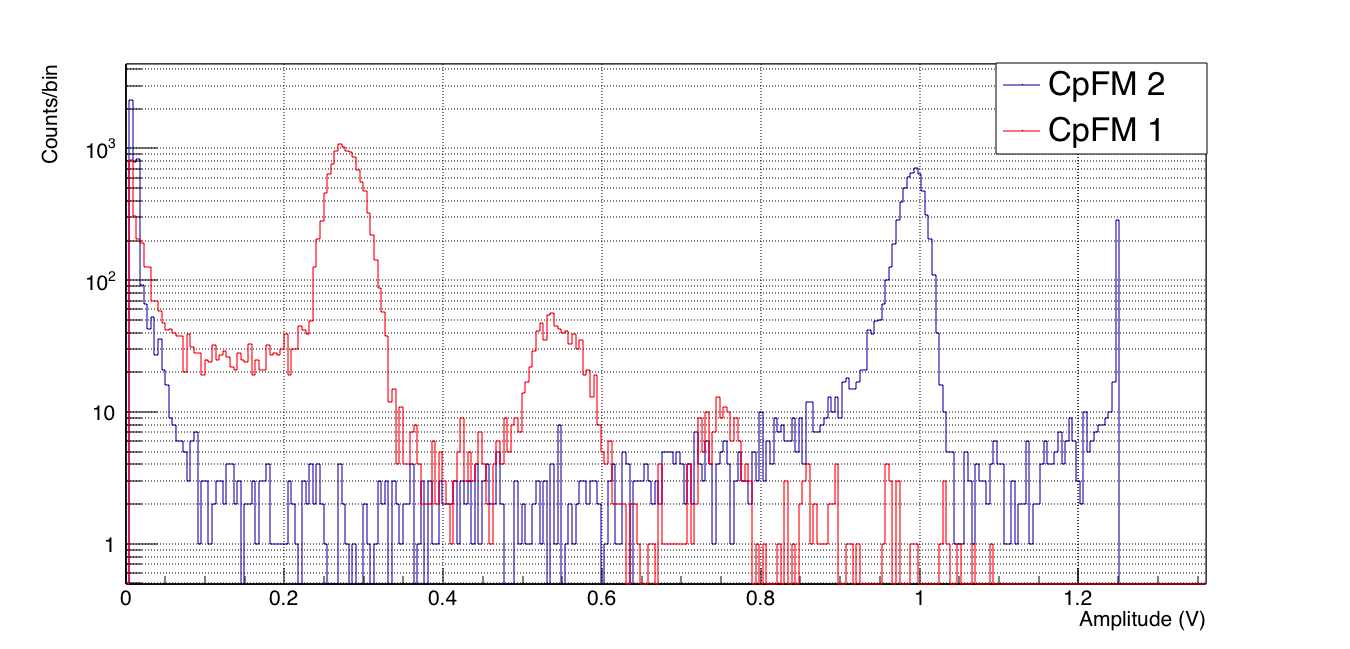}}
\subfigure[\label{CpFM1_ion}]{\includegraphics[width=0.68\textwidth]{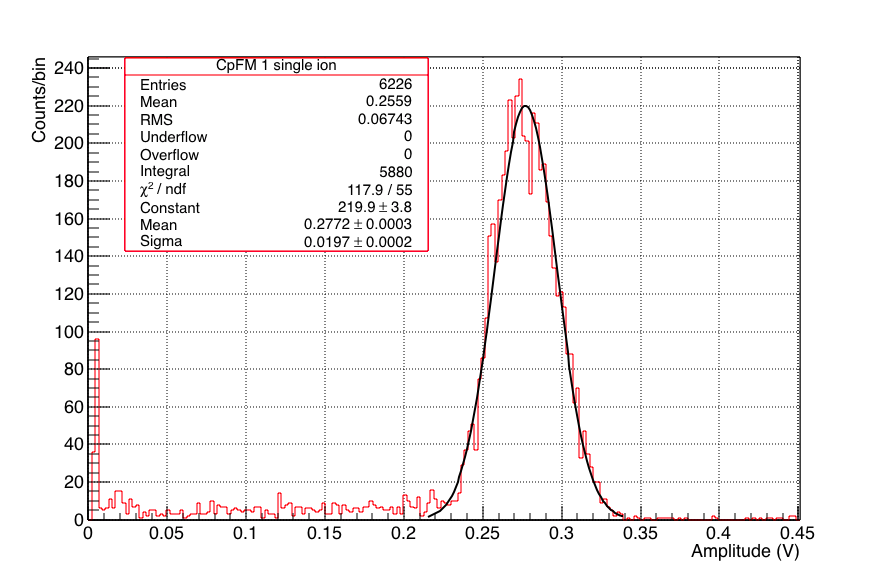}}
\caption{(a) Amplitude distribution of the CpFM 1 (red line) and the CpFM 2 (blue line) signals corresponding to the whole data taking. Pb ion run, November 2016. In the CpFM 1 distribution the two well defined peak correspond to a single ion and double-ion signal respectively. In the CpFM 2 distribution it is visible only the single ion peak because the electronics saturation occurs before the value at which the double-ion peak is expected. (b) CpFM 1 amplitude distribution corresponding to the single ion peak. The remotion of the background events (mainly due to ion fragments) has been performed through a cut on the signal shape width. November 2016. Pb-ion run.}
\end{center}
\end{figure}
 In Fig.\,\ref{ion_distribution} the amplitude distributions of the CpFM channels are shown (CpFM 1 red line, CpFM 2 blue line) where only a rough requirement on the amplitude ($>$6\,mV) to cut the electronic noise has been applied. In the CpFM 1 distribution a three peak structure is present corresponding to 1, 2 and, just hinted, 3 ions. In the CpFM 2 distribution only one peak appears together with the electronic saturation occurring at 1.25\,V (no offset was used). This is explained by different calibration factors. Fitting with a Gaussian function the single-ion peak as shown in Fig.\,\ref{CpFM1_ion} for CpFM 1 and inverting the Eq. \ref{recal}, the calibration factors for the CpFM channels are:
${y}_{CpFM1}= 0.066\pm 0.001\,(ph.e/p) $ and ${y}_{CpFM2}= 0.1862\pm 0.0004\,(ph.e/p) $ for the CpFM 1 and the CpFM 2 respectively.
The CpFM 2, devoted to the background measurement, results about 3 times more efficient than the CpFM 1; both the efficiency values are also lower with respect to the one of the CpFM prototype tested at the BTF. This was due to a problem during the installation investigated and solved during the winter SPS shut-down \cite{NATOCHII201815}.
The efficiency ($\epsilon$) of this version of the detector is well described by an upper cumulative distribution function of a Binomial distribution $B(k, n, p)$, being $n$ the real number of incoming protons to be detect, $k$ the total number of photoelectrons produced by the $n$ protons and p the single proton efficiency of the CpFM: 
\begin{equation}
\label{binomial}
\epsilon = Q(1, n, p) =\sum_{t=1}^n  B(1, n, p)
\end{equation}
Using this model with $p={y}_{CpFM}$, the expected number of photoelectrons ($k$) produced per $n$ protons can be determined. Multiplying $k$ by the value in mV of one ph.e (corresponding to the peak in Fig.\,\ref{single_photoelectron} at 1050 V) and comparing the result with the amplitude of the electronic noise ( $<6$\,mV at 1050 V) is thus possible to assert that the CpFM is effective in counting the incoming protons if  $n>6$ for the CpFM 1 and  $n>2$ for the CpFM 2.

 \section{Commissioning and operations}
In this section the most common operations in which the detector is involved are described. During the commissioning phase they were also used to validate the functionality of the detector, allowing the measurement of some well know channeled beam and crystal characteristics.\\
The crystal assisted collimation prototype is composed essentially by the crystal bending in the horizontal plane and the absorber (Fig.\,\ref{ua9setup}). The crystal acts as a primary collimator deflecting the particles of the halo toward the absorber ($\simeq$65\,m downstream the crystal area) which has the right phase advance to intercept them and represents the secondary target. The CpFM is placed downstream to the crystal ($\simeq$60\,m) and intercepts and counts the particles before they are absorbed. More details about the UA9 setup and procedure to test the crystal collimation can be found in \cite{MontesanoIPAC}.
\subsection{Standard operation: angular scan}
The angular scan of the crystal is the UA9 standard procedure through which the channeling orientation of the crystal is identified and the experimental setup becomes operational.
It consists in gradually varying the orientation of the crystal with respect to the beam axis, searching for the \textit{planar channeling} and \textit{volume reflection}\footnote{it is the coherent interaction that a charged particle experiences in bent crystal when the its angle with respect to the crystal planes is bigger than the critical angle for channeling and smaller than the bending angle of the crystal} regions \cite{biryukov}, while the crystal transverse position is kept constant. When the optimal channeling orientation is reached, the number of inelastic interactions at the crystal is at minimum and the number of deflected particles is at its maximum. Consequently the loss rate measured by the BLMs close to the crystal should show a suppression while in the CpFM signal rate a maximum should appear.\\
\begin{figure}[htbp!]
\begin{center}
\includegraphics[width=0.8\textwidth]{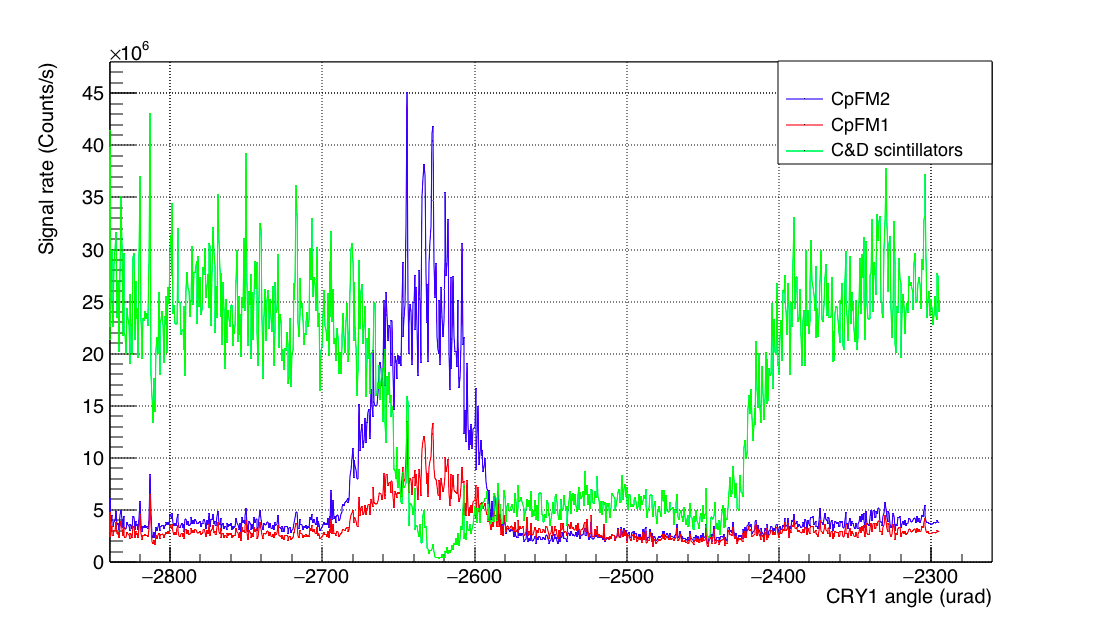}
\caption{Angular scan of UA9 crystal 1 (bending angle = 176 $\mu$rad) during the proton MD in October 2016 (scan speed is $0.5\,\mu$rad/s). In the optimal channeling orientation the CpFM 2 (blue line) signal rate is $\sim$3 times bigger than the signal rate of the channel 1 (red line). This is due to the different efficiency as shown in Sec. \ref{calibration}. The BLMs signal rate corresponds to the coincident signal rate of a pair of plastic scintillators located at the crystal position.}
\label{Angular_scan_prot}
\end{center}
\end{figure}
In Fig.\,\ref{Angular_scan_prot} the angular scan of the UA9 crystal 1 during a proton run is shown. It is displayed both by the BLMs and the CpFM (CpFM position is such that both the bars intercept the whole channeled beam when the crystal is in the optimal channeling position). The first and the last angular regions (angle$<$ -2700\,$\mu$rad and angle$>$-2400\,$\mu$rad) correspond to the \textit{amorphous} orientation. As expected, here the loss rate registered by the BLMs is maximum while the CpFM signal rate is minimum. After the first amorphous region, the channeling peak appears (around angle of - 2620\,$\mu$rad) as a maximum in the CpFM rate and as minimum in the BLMs rate. Just after, only in the BLMs signal, the \textit{volume reflection} area is clearly visible. In this angular region the particles experience a deflection to the opposite side with respect to the planar channeling deflection. For this reason \textit{volume reflection} is not detectable by the CpFM except as a slight reduction in the background counts. 
Although the BLMs rate profile is an effective instrument for the estimation of the best channeling orientation, it is based on beam losses and is generally less sensitive than the CpFM rate profile which on the contrary measures directly the presence of channeled protons.  
\subsection{Standard operation: Linear scan}
 \begin{figure*}[htbp!!]
 \centering
\subfigure[\label{tesivale}]{\includegraphics[width=0.3\textwidth]{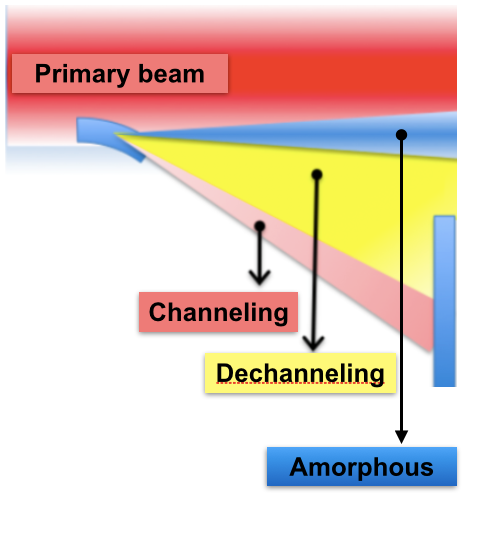}}
\subfigure[\label{linear_scan}]{\includegraphics[width=0.68\textwidth]{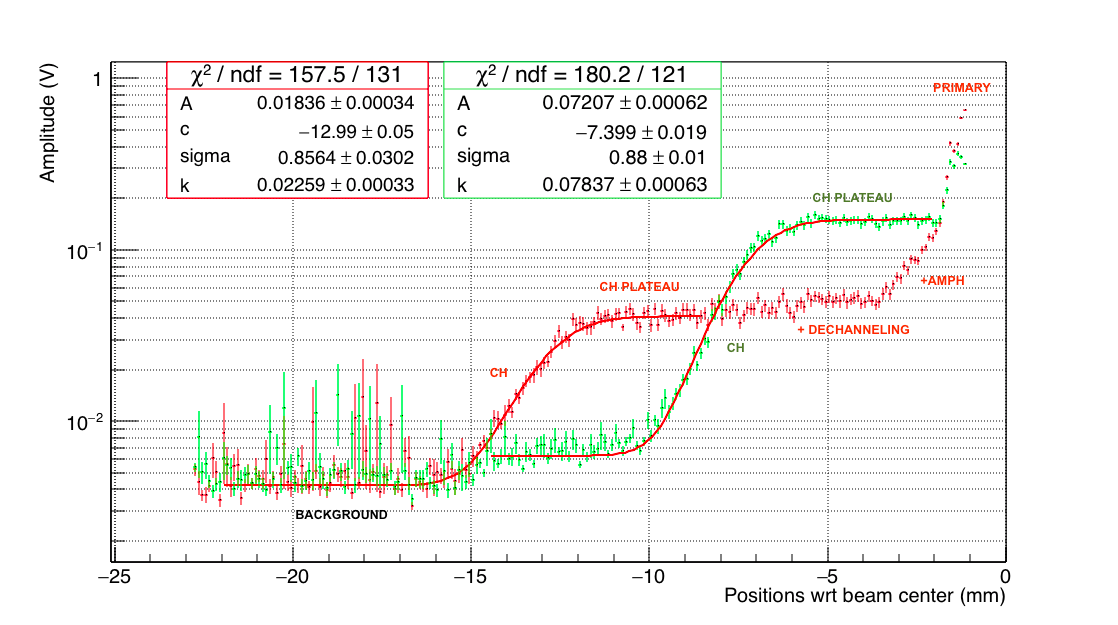}}
\caption{(a) Conceptual sketch of the particle components that the CpFM progressively intercepts. (b) CpFM Linear scan with the UA9 crystal 4 during the proton beam run in October 2015. The particle interactions with the crystal corresponding to the different regions of the linear scan are also pointed out. In the channeling region, the plots correspond to the integrated measurement of the channeled beam horizontal profile and they can be fitted with an error function. The x axis corresponds to the CpFM 1 distance from the center of the primary beam.}
\label{regions}
\end{figure*}
The CpFM linear scan is the standard procedure needed to identify the CpFM position with respect to the primary and the channeled beam. 
 A fast linear scan (linear motor speed $\simeq100$\,$\mu m/s$) is performed at the very beginning of the operations, when the crystal has not been yet positioned in the channeling orientation, to align the CpFM with respect to the beam core. 
 The CpFM is gradually inserted until the aperture of the detector corresponds to the beam aperture and a sharp spike is observed in the rate monitor of the WaveCatcher. \\
A slower linear scan (linear motor speed $\simeq10$\,$\mu$m/s) is instead performed when the crystal is already correctly oriented to extract the halo particles. Fig.\,\ref{linear_scan} shows an example of slow scan for the proton run of October 2015. The plot shows the average amplitude value of both the channels of the CpFM as a function of the distance of the CpFM 1 with respect to the beam core. As expected, the CpFM 1 bar starts to intercept the channeled protons first. The signal amplitude increases indeed on this channel.  Around 10\,mm from the beam core CpFM 1 detects the whole channeled beam and enters in the so-called \textit{channeling plateau} where the amplitude signal is consequently constant. Moving the detector further inside the signal increases: initially it increases with a slow rate since the first bar intercepts the \textit{dechanneled}\footnote{particles initially channeled which escape from the channeling mode due to Coulomb scattering on atoms. These particles are less bent than the channeled ones, depending on how long they are kept in the channeling state before they lose it} particles, and then exponentially, when also particles interacting with the crystal as it was in the \textit{amorphous} orientation are intercepted \ref{tesivale}. These particles underwent multiple Coulomb scattering inside the crystal and thus they are displaced with respect to the beam core by few $\mu$rad. The CpFM 2 detects the same particle flux but with a spatial displacement of 5\,mm with respect to the CpFM 1, being exactly 5\,mm the distance between the two bars edges.
\subsubsection{Channeled beam profile}
In the \textit{channeleing plateau}, the linear scan shown in Fig.\,\ref{linear_scan} basically corresponds to integrate the channeled beam profile in the horizontal plane. Therefore it can be fitted with an error function:
\begin{equation}
\label{erf}
erf(x)= A\cdot \frac{1}{\sigma\sqrt{2\pi}}\int_0^x e^{-\frac{(t-c)^2}{2\sigma^2}} dt +K
\end{equation}
Where $\sigma$ is the standard deviation of the Gaussian beam profile, $c$ is the center of the channeled beam with respect to the primary one and $A$ and $K$ are constants related to the \textit{channeling plateau} value and the background value.\\
In Fig.\,\ref{linear_scan} CpFM 1 and CpFM 2 linear scan profiles of Fig.\,\ref{linear_scan} have been fitted with the error function described in Eq. \ref{erf}. 
From the results of the fits the channeled beam size ($\sigma$) at the position of the CpFM is obtained as well as some informations confirming the functionality of the detector: 
both the CpFM 1 and CpFM 2 measure compatible values of the channeled beam standard deviation ($\sigma$) and, as expected, the difference between the channeled beam center (c) measured by the CpFM 1 and the CpFM 2 is compatible with the design distance between CpFM 1 and CpFM 2 bar edges.
\subsubsection{Crystal bending angle and angular spread of the channeled beam at the crystal position}
The results of the fits performed on the integrated beam profiles in Fig.\,\ref{linear_scan} provide two additional functionality tests of the detector allowing to derive channeled beam and crystal characteristics already well known.

In particular the center ($c$) of the channeled beam can be used to determine the value of the bending angle $\theta_{bend}$ of the crystal. This represents a non perturbative method to measure in-situ the crystal bending angle, alternative to the linear scan of the LHC-type collimator used in the past \cite{Previtali}. The CpFM, unlike the collimator (1\,m of carbon fiber composite), is indeed almost transparent to the channeled protons which produce Cherenkov light losing a negligible amount of their energy. \\
$\theta_{bend}$ is derived calculating the equivalent crystal kick $\theta_k$ at the CpFM position along the ring. The latter corresponds to the angular kick that a particle should receive by the crystal to be horizontally displaced by x with respect to the beam core at the CpFM position. It is derived applying the general transfer matrix to the phase-space coordinates of the particle at the crystal position $(x_{0}, x'_{0})_{cry}$ to get the new coordinate at the CpFM position the CpFM $(x, x' )_{CpFM}$:
\begin{equation}
\label{teta_k}
\theta_{k}=\frac{x_{CpFM}-\sqrt{\frac{\beta_{CpFM}}{\beta_{cry}}} x_{cry}cos\Delta\phi} {\sqrt{\beta_{cry}\beta_{CpFM}} sin\Delta\phi}
\end{equation}
being $\beta_{CpFM}$ and $\beta_{cry}$ the Twiss parameter at the CpFM and at the crystal location respectively and $\Delta\phi$ the phase advance between the crystal and the CpFM. These values are tabulated for the SPS machine \cite{UA9} ($\beta_{cry}=87.1154$\,m, $\beta_{CpFM}=69.1920$\,m, $\Delta\phi/2\pi=0.23244$). More details about this mathematical procedure can be found in \cite{Addesa:2661725}.
When $x_{CpFM} =c$, the equivalent kick $\theta_{k}$ corresponds to the bending angle of the crystal $\theta_{bend}$.
Using the value of $c$ as extrapolated by the fit (see Fig.\,\ref{linear_scan}) it is now possible to determine $\theta_{bend}$ corresponding to the crystal used during the scan: $\theta_{bend}= (167\pm 6)\,\mu rad$.
Its bending angle was previously measured by means of interferometric technics (Veeco NT1100) and resulted to be 176\,$\mu$rad. 
The slight discrepancy with the CpFM indirect measurement of the bending angle could depend on the imprecise evaluation of the primary beam center during the CpFM alignment procedure, not accounted in the error.\\
Using the value above and the value of the $\sigma$ of the channeled beam obtained by the fit shown in Fig.\,\ref{linear_scan}, it is also possible to extrapolate the angular spread of the particles exiting the crystal.
It can be derived subtracting the equivalent kick for $x_{CpFM}$ = c$\pm\sigma$ from $\theta_{bend}$, corresponding to the equivalent kick calculated in the center $c$ of the channeled beam:
\begin{equation}
\label{spread}
\theta_{spread}=[\theta_{k}]_{c \pm\sigma} \mp\theta_{bend}
\end{equation}
applying the Eq. \ref{spread} to the fit results in Fig.\,\ref{linear_scan}, the angular spread has been evaluated to be:  $\theta_{spread}=(12.8\pm1.3)\,\mu rad$.\\
The angular spread at the exit of the crystal is directly connected to the critical angle value which defines the angular acceptance of the channeled particles at the entrance of the crystal. Therefore the angular spread should be comparable with respect to the critical angle. From theory \cite{biryukov}, for 270 GeV protons in Si $\theta_c$ is 12.2 $\mu$rad\footnote{for Si planes orientation (110)}. It can be then asserted that the angular spread derived by the fit results and the critical angle computed from the theory are well comparable. 
\subsection{Flux measurement}
The detector position optimal for the measurement of the channeled particle flux is the channeling plateau for the inner bar and the background position for the external one. In this position, the CpFM 1 bar intercepts the whole channeled beam while the CpFM 2 bar measures only the background. If both the channels have equal efficiency, a more accurate flux measurement is possible in this location since the background signal can be subtracted from the channeled beam signal on event by event basis. 
During the commissioning of the detector, both CpFM channels were tested in their own channeling plateau positions to compare the results of the flux measurement while both bars intercept the whole channeled beam.
 \begin{figure}[htbp!!]
 \centering
\subfigure[\label{Number_protons_new_calibration}]{\includegraphics[width=0.68\textwidth]{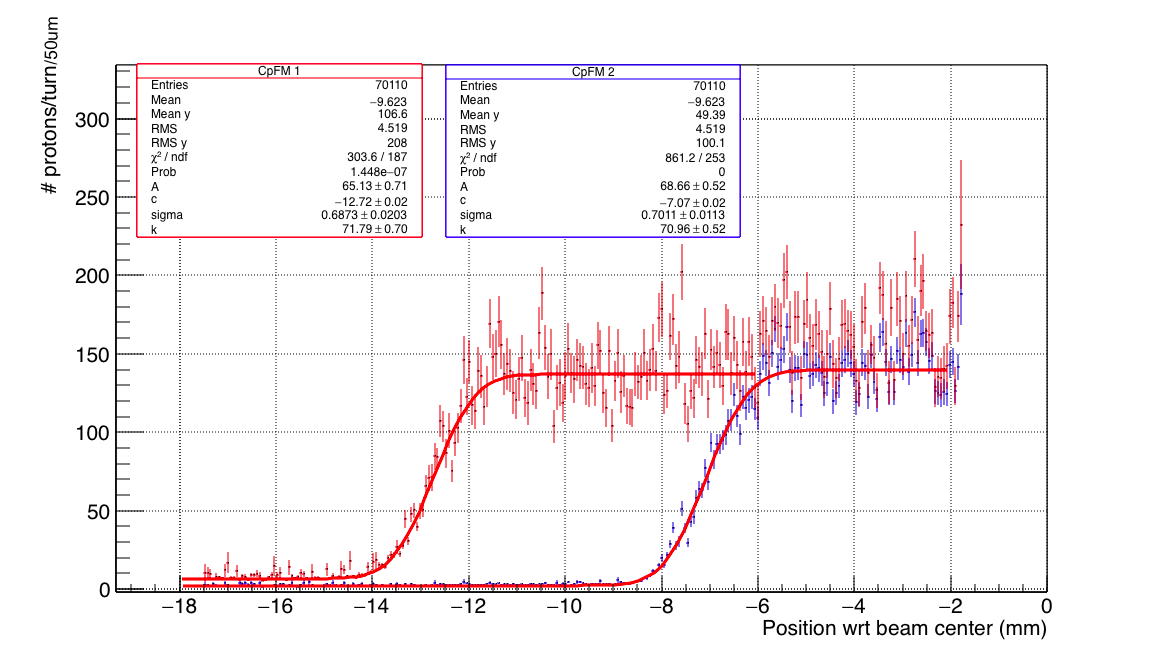}}
\subfigure[\label{ch_plateau_new_4sigma}]{\includegraphics[width=0.68\textwidth]{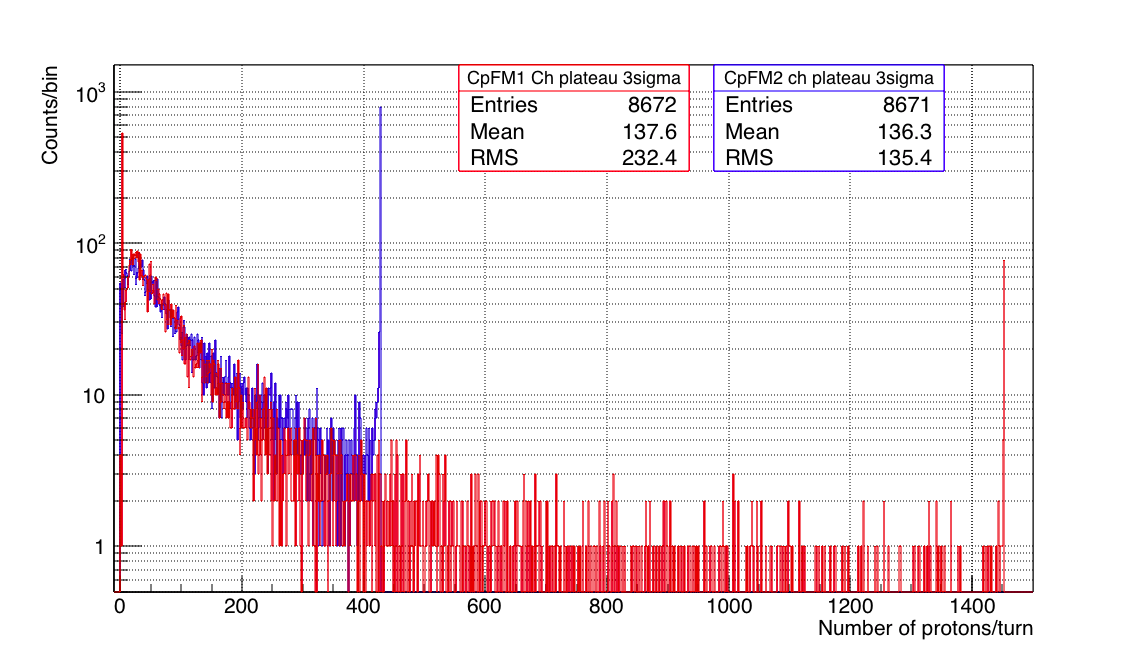}}
\caption{(a) CpFM linear scan performed during October 2016 with 270 GeV protons and crystal 1 at 3$\sigma$ from the main beam. It represents the average number of protons extracted per turn as a function of the CpFM position. Each bin corresponds to 50 $\mu$m (5 sec of data acquisition, namely 215 trigger events). The calibration factors have been applied and the CpFM 1 and the CpFm 2 channels are equalized. The number of extracted protons detected is compatible with the BCTDC measurements. (b) Distribution of the number of protons counted by the CpFM 1 (red) and CpFM 2 (blue) per turn when the new calibration factors are introduced. Both the distributions are related to data collected during 200 seconds in the channeling plateau of the CpFM 1 (-11 to -9 mm from beam center) and CpFM 2 (-5 to -3 mm from beam center) as displayed in Fig.\,\ref{Number_protons_new_calibration}}
\end{figure}
In Fig.\,\ref{Number_protons_new_calibration} the linear scan profile performed in October 2016 with 270 GeV protons is shown.
It is a similar plot as the one shown in Fig.\,\ref{linear_scan}, but in this case the amplitude value is divided by the calibration factor, as found in Sec. \ref{calibration} and expressed in mV/proton, to display the number of channeled protons extracted per turn.
In Fig.\,\ref{ch_plateau_new_4sigma} the distributions of number of protons related to the channeling plateau regions of CpFM 1 and CpFM 2 are shown. No event selection has been performed since a pedestal event is caused either by the physical absence of channeled particles or by the inefficiency of the detector. In fact, the absence of channeled particles could be connected to orbit instabilities, beam halo dynamics or to the inefficiency of the crystal-extraction system, with the latter under study by the CpFM detector. The inefficiency of the detector is already taken into account by the calibration factors.
In Fig.\,\ref{ch_plateau_new_4sigma} pedestal events are much more abundant in the CpFM 1 distribution than in the CpFM 2 distribution. In this case the pedestal events are mostly due to the inefficiency of the CpFM 1 bar, a factor of 3 worse than the efficiency of the CpFM 2 bar. Indeed, when the number of extracted particles is low ($<6$), the CpFM 1 can not discriminate extracted protons by the electronic noise. 
Both the channels count in average approximately the same number of protons per turn; the slight difference being due to the saturation of the electronics which occurs for the two channels at a different number of protons per pulse and with different percentages (1\% of the entries for CpFM 1, 10\% for CpFM2).
In order to validate these results the flux extracted from the halo beam was estimated by the Beam Current Transformer (BCTDC \cite{Jones:1982418}) installed in the SPS. BCTDC integrates the beam current along an SPS revolution (23\,$\mu sec$) measuring the total charge circulating in the machine.  The time derivative of the BCTDC corresponds to the total particle flux leaving the machine; since the crystal acts as a primary target in the machine, the beam intensity variation can be assumed to be mainly caused by it and hence corresponding to the flux detected by the CpFM. This is an approximation as other minor losses can occur in the machine.
With the typical fluxes extracted by the crystal ($10^5 - 10^7 $\,p/s) BCTDC measurements are only reliable when averaged over time intervals of several seconds. 
The extracted flux estimated by the BCTDC in the time interval related to Fig.\,\ref{ch_plateau_new_4sigma} is $179\pm42$ protons per turn. This value has to be considered in good agreement with the values of the flux measured by CpFM 1 and CpFM 2 remembering that the BCTDC measurement represents the upper limit estimation of the crystal-extracted flux.\\
Finally, particular attention has to be paid to the shape of the distributions in Fig.\,\ref{ch_plateau_new_4sigma}. They are not Gaussian. For such a high fluxes, this can not depend on the detector resolution, at least for the CpFM 2 channel which has the better efficiency.
This can be demonstrated deriving the CpFM 2 resolution for an incident and constant flux of 180 protons per turn, as measured by the BCTDC. 
From the single ion distributions in Fig.\,\ref{calibration}, the resolution with respect to a single incident lead ion can be computed.
The CpFM 2 resolution for 180 protons is easily derived scaling the ion resolution by the factor $\sqrt(6724/180)$. It corresponds to  9\%.
Thus, if the number of protons extracted by the crystal was constant and equal to 180 (distributed according a Gaussian distribution centered in 180), the CpFM 2 signal would be characterized by a narrower peak having 15 protons $\sigma$. 
The beam extracted by the crystal is therefore not constant on the time scale probed by the CpFM. 
There are several possible reasons for this: the diffusion dynamics of the halo beam, goniometer instabilities or orbit instabilities. 
The CpFM detector offers an interesting chance to address this issue at the $\mu$s scale but the current data acquisition electronics of the detector represent a limit. The CpFM detector is indeed able to accept only 1/1000 SPS trigger, since the data acquisition electronics is not fast enough ($<1$\,kHz). A faster electronics, matching the revolution frequency of the machine, could strengthen the detector capability in studying the impact of the listed factors on the crystal halo extraction.
\section{ CpFM 2.0: in-situ calibration with Xenon ions and first case study}
During the winter shut-down of 2016, the layout of the CpFM detector was modified. In order to improve the detector efficiency the fiber bundles were removed. Only one PMT was directly coupled to the viewport and in such a way that the transversal cross section of both bars is covered.
\begin{figure}[htbp!!!]
\centering
\includegraphics[width=0.6\textwidth]{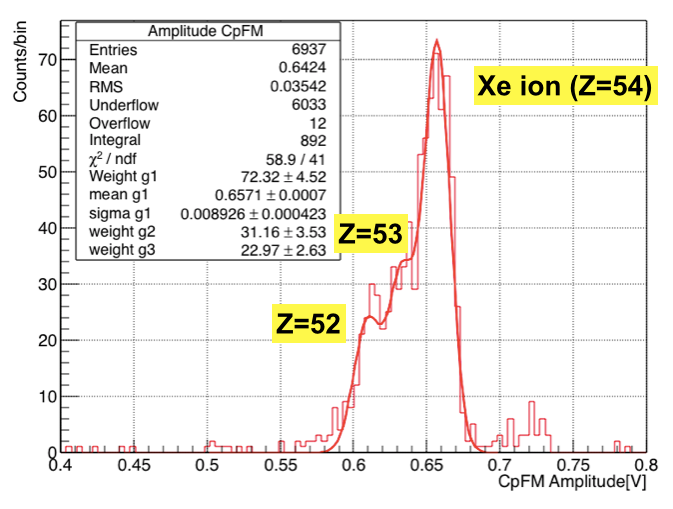}
\caption{Xenon ions amplitude distribution. Beside the ion peak are also visible two peaks associated to ion fragments with charge $Z=53$ and $Z=52$. The fit function is the sum of three Gaussian functions. Only Gaussian related to the Xe ions has 3 free parameters. For the other two, the only free parameter is the weight (g) while the mean and the sigma values are fixed by the mean and the sigma values of the Xe ion Gaussian, being derived by a scale factor dependent only on the charge Z (for more detail see  \cite{Addesa:2661725}}
\label{Xenon}
\end{figure}
The fibers bundle was indeed responsible for a reduction factor of 10 in the light yield per proton. Moreover CpFM 1 and CpFM 2 bars were inverted, the latter being better polished and thus more efficient.
In December 2017, the second version of the CpFM detector was calibrated with a 270 GeV/charge Xenon (Xe, $Z=54$) ion beam using the same procedure described in Sec. \ref{calibration}. In Fig.\,\ref{Xenon} the signal amplitude distribution focused on the Xenon ion events is shown. It is referred to a data taking performed during a crystal angular scan when only the inner bar intercepted the channeled beam. Beside the main peak due to Xe ions, two other structures appear. They are associated with fragments of charge Z=53 and Z=52 respectively, produced when the crystal is not in the optimal channeling orientation. Using the results of the fit, the new calibration factor is derived:${y}_{CpFM}= 2.1\pm 0.2\,(ph.e/p)$.
As expected it results improved by a factor of 10 with respect the old version of the detector.\\
 \begin{figure}[htbp!!]
 \centering
\subfigure[\label{sketch_xenon}]{\includegraphics[width=0.44\textwidth]{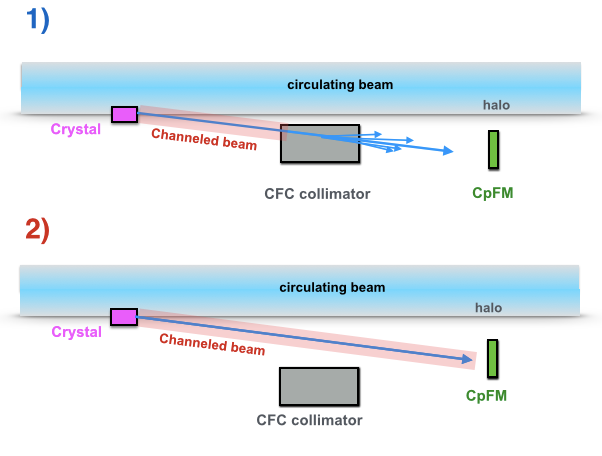}}
\subfigure[\label{distribution_xenon}]{\includegraphics[width=0.55\textwidth]{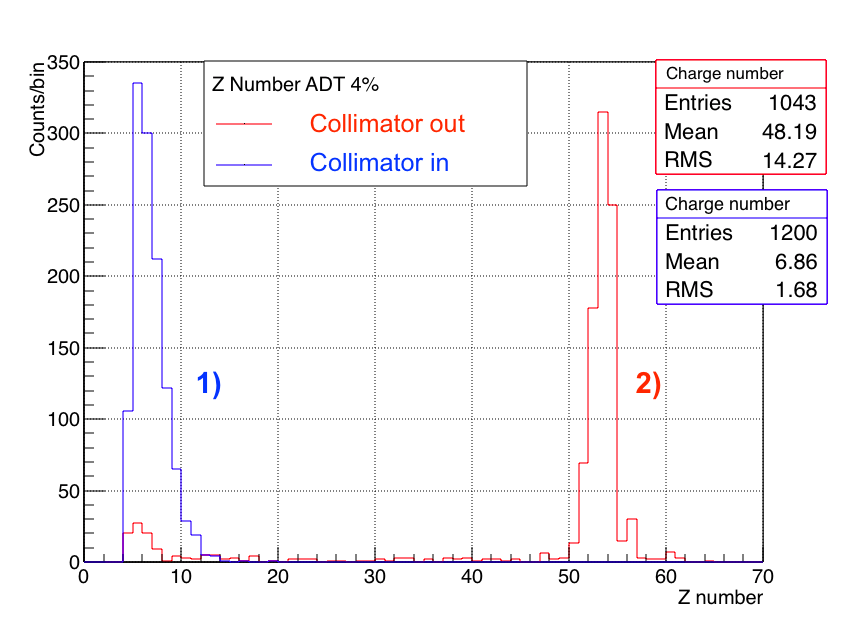}}
\caption{(a) Conceptual sketch of the crystal-collimation setup when the CFC collimator is inserted (1) and retracted (2). (b) Signal charge distribution referred to setup 1) and 2). Only a pedestal cut (signal amplitude $<5$\,mV has been applied.}
\label{collimation}
\end{figure}
After the calibration, the detector was used to observe the particle population exiting 1\,m long CFC (Carbon Fiber Composite) LHC-like collimator when Xenon ions are deflected onto it.  
The collimator is part of the UA9 crystal-assisted collimation setup. It is located downstream the crystals region and about 17\,m upstream the CpFM. During the case study, the Xe ions were channeled and deflected onto the collimator. The CpFM were thus inserted to detect the channeled beam after the passage through the collimator. The measurement was repeated retracting the collimator. The results are shown in Fig.\,\ref{collimation}
The CpFM successfully discriminated the low-Z particle population (mostly $Z<$6) resulting from the fragmentation of Xe ions inside the collimator from the Xe ions themselves. 
\section{Conclusion}
The CpFM detector has been developed in the frame of the UA9 experiment with the aim to monitor and characterize channeled hadron beams directly inside the beam pipe vacuum. It consists of fused silica fingers which intercept the particles deflected by the crystal and generate Cherenkov light.
The CpFM is installed in the UA9 crystal collimation setup in the SPS tunnel since 2015. It has been successfully commissioned with different beam modes and with proton and ion beams and it is now fully integrated in the beam diagnostic of the experiment, providing the channeled beam flux measurement and being part of the angular alignment procedures of bent crystals. It is able to provide the channeled beam horizontal profile and the measurement of crystal-extracted flux with a relative resolution of about 11$\%$ for 100 protons/bunch. In order to improve the detector resolution for lower fluxes a new radiator geometry is under study.\\
The CpFM offers an interesting chance to investigate halo beam dynamics in the crystal-extraction process at the $\mu s$ scale but the current data acquisition represents a limit.  A faster data acquisition system, matching the revolution frequency of the SPS machine (43kHz), could strengthen the detector capability to study and characterize crystal-extraction in a circular machine.


\bibliography{mybibfile}

\end{document}